\documentclass[10pt,twocolumn,letterpaper]{article}

\usepackage{cvpr}
\usepackage{times}
\usepackage{epsfig}
\usepackage{graphicx}
\usepackage{amsmath}
\usepackage{amssymb}
\usepackage{subfig}
\usepackage{booktabs}
\usepackage{caption}
\usepackage[pagebackref=true,breaklinks=true,letterpaper=true,colorlinks,bookmarks=false]{hyperref}

\cvprfinalcopy % *** Uncomment this line for the final submission

\def\cvprPaperID{30} % *** Enter the CVPR Paper ID here

\ifcvprfinal\fi
\begin{document}

%%%%%%%%% TITLE
\title{Farm Parcel Delineation Using Spatio-temporal Convolutional Networks}

\author{Han Lin Aung\\
Department of Computer Science\\
Stanford University\\
%Stanford, CA 94309, USA\\
{\tt\small hanlaung@cs.stanford.edu}
\and
Burak Uzkent\\
Department of Computer Science\\
Stanford University\\
%Stanford, CA 94309, USA\\
{\tt\small buzkent@cs.stanford.edu}
\and
Marshall Burke\\
Department of Earth Systems Science\\
Stanford University\\
%Stanford, CA 94309, USA\\
{\tt\small mburke@stanford.edu}
\and 
David Lobell\\
Department of Earth Systems Science\\
Stanford University\\
%Stanford, CA 94309, USA\\
{\tt\small dlobell@stanford.edu}
\and
Stefano Ermon\\
Department of Computer Science\\
Stanford University\\
%Stanford, CA 94309, USA\\
{\tt\small ermon@cs.stanford.edu}
}

\maketitle
%\thispagestyle{empty}

%%%%%%%%% ABSTRACT
\begin{abstract}
 Farm parcel delineation (delineation of boundaries of farmland parcels/segmentation of farmland areas) provides cadastral data that is important in developing and managing climate change policies. Specifically, farm parcel delineation informs applications in downstream governmental policies of land allocation, irrigation, fertilization, greenhouse gases (GHG's), etc. This data can also be useful for the agricultural insurance sector for assessing compensations following damages associated with extreme weather events - a growing trend related to climate change~\cite{childress2014linking}. Using satellite imaging can be a scalable and cost-effective manner to perform the task of farm parcel delineation to collect this valuable data. In this paper, we break down this task using satellite imaging into two approaches: 1) Segmentation of parcel boundaries, and 2) Segmentation of parcel areas. We implemented variations of U-Nets, one of which takes into account temporal information, which achieved the best results on our dataset on farm parcels in France in 2017.

\end{abstract}

%%%%%%%%% BODY TEXT
\section{Introduction}
Farm parcel delineation has been a highly manual task before the use of computer vision and machine learning, incurring high costs and time for those who are labeling the data through `theodolites, total stations, and GPS'~\cite{van2002use}. Beyond costs, the cadastral information retrieved from delineation of farm parcels is particularly important in forming climate change policies for mitigation and adaptation. More specifically, such information is important in developing and managing incentive plans on environmental and climate change, allocation of water and irrigation, and policies around agricultural insurance for catastrophe. Overall, according to World Bank, land policies, specifically for farm parcels, are considered to be highly intertwined with climate change~\cite{childress2014linking}. Hence, we believe that automating farm parcel delineation in a scalable manner can cut down costs and time by replacing the otherwise manual process of collecting parcel-level information. Doing so can help inform relevant policy makers and stakeholders on real-time cadastral data very rapidly. 

On the technical front, deep convolutional neural networks (CNNs) have been successfully applied to different computer vision tasks including image recognition~\cite{krizhevsky2012imagenet,he2016deep,simonyan2014very}, object detection~\cite{uzkent2020efficient,lin2014microsoft,ren2015faster,redmon2017yolov2}, and object tracking~\cite{kristan2017visual,bertinetto2016fully,uzkent2017aerial,uzkent2018enkcf}. Following this trend, we investigate its application in delineating farm parcel boundaries/areas using satellite imagery. Combination of CNNs with satellite images is a scalable method that automates the otherwise manual process~\cite{sheehan2019predicting,uzkent2019learning}. We further take into account temporal data to improve results over the vanilla approach using a single image for our farm parcel delineation task. 
\begin{figure*}[t!]%
\centering
\begin{tabular}{c}
{{\includegraphics[width=0.185\linewidth]{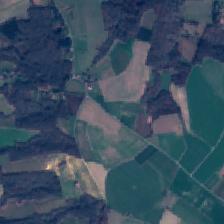} } {\includegraphics[width=0.185\linewidth]{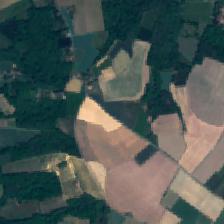} } {\includegraphics[width=0.185\linewidth]{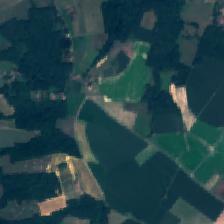} }
{\includegraphics[width=0.185\linewidth]{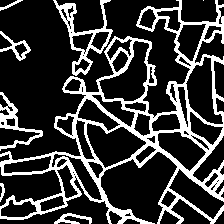} }
{\includegraphics[width=0.185\linewidth]{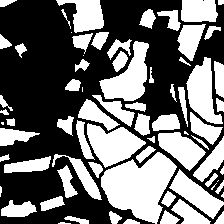} }}\\
{{\includegraphics[width=0.185\linewidth]{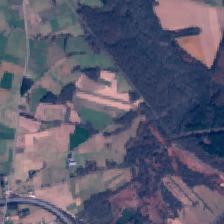} } {\includegraphics[width=0.185\linewidth]{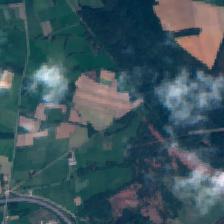} } {\includegraphics[width=0.185\linewidth]{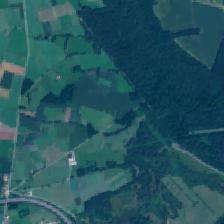} }
{\includegraphics[width=0.185\linewidth]{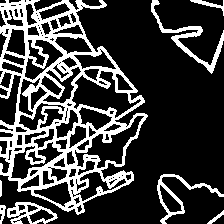} }
{\includegraphics[width=0.185\linewidth]{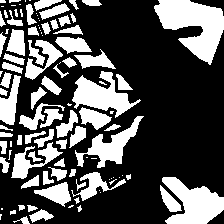} }}\\ 
{{\includegraphics[width=0.185\linewidth]{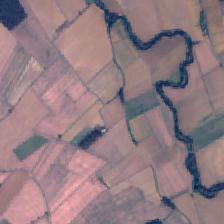} } {\includegraphics[width=0.185\linewidth]{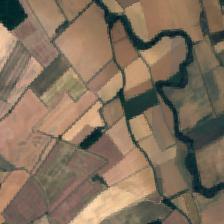} } {\includegraphics[width=0.185\linewidth]{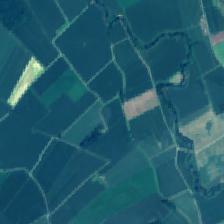} }
{\includegraphics[width=0.185\linewidth]{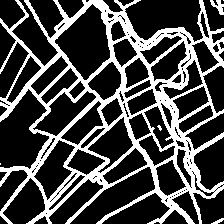} }
{\includegraphics[width=0.185\linewidth]{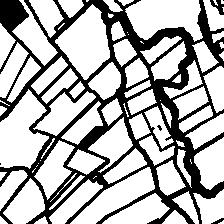} }}\\
{{\includegraphics[width=0.185\linewidth]{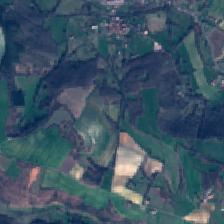} } {\includegraphics[width=0.185\linewidth]{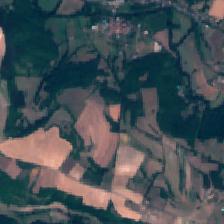} } {\includegraphics[width=0.185\linewidth]{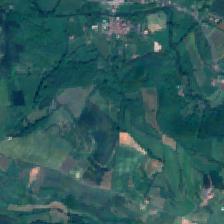} }
{\includegraphics[width=0.185\linewidth]{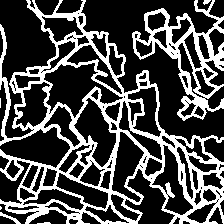} }
{\includegraphics[width=0.185\linewidth]{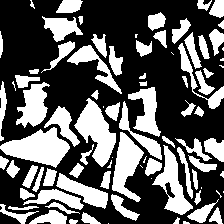} }}

%, \\ 
% \subfloat[]{{\includegraphics[width=1cm]{jan_march_4.jpeg} }, {\includegraphics[width=1cm]{apr_jun_4.jpeg} }, {\includegraphics[width=1cm]{jul_sept_4.jpeg} },
% {\includegraphics[width=1cm]{out_4.png} },
% {\includegraphics[width=1cm]{out_4_filled.png} }
\end{tabular}
    % \centering
    %   \subfloat[Ground truth (satellite image overlay) ]{{\includegraphics[width=2.7cm]{ground_truth.png} }, {\includegraphics[width=2.7cm]{Predict_border_0.png} }, {\includegraphics[width=2.7cm]{Predict_area.png} }}%
    % \newline
    % \centering
    % \subfloat[Ground truth (satellite image overlay) ]{{\includegraphics[width=2.7cm]{ground_truth.png} }, {\includegraphics[width=2.7cm]{Predict_border_0.png} }, {\includegraphics[width=2.7cm]{Predict_area.png} }}%
    % \newline 
    %   \subfloat[Ground truth (satellite image overlay) ]{{\includegraphics[width=2.7cm]{ground_truth.png} }, {\includegraphics[width=2.7cm]{Predict_border_0.png} }, {\includegraphics[width=2.7cm]{Predict_area.png} }}%
    \caption{Each row is an instance from the dataset. From left to right, the first 3 images are satellite images from January-March, April-June, July-September respectively. The final 2 images are the binary image masks of boundaries and areas respectively.}%
    \label{fig:example_data_instance}%
\end{figure*}
\section{Related Work}
%Apart from the improved performance deep learning provides over more traditional techniques, deep learning has also been able to replace the manual process of using multiple feature extraction algorithms to retrieve hand crafted features. Furthermore,%

In recent years, deep learning has become very popular in computer vision tasks due to its incredible success. Known as a universal learning approach, the same deep neural network architecture can be applied with great success over multiple domains~\cite{deep_learning_survey}. One particular domain in computer vision that deep learning has shown state-of-the-art results is on image segmentation tasks. 

There are two general Convolutional Neural Network (CNN) architectures often used in image segmentation: patch-based networks and Fully Convolutional Networks (FCNs). The patch-based models~\cite{springenberg2014striving} receive fixed-size patches based on a central pixel as inputs to make a prediction. Hence, to make predictions on a per-pixel level, these models receive patches for each of the pixels of each input image. However, using patch-based models to provide dense predictions on a per-pixel level is highly computationally expensive~\cite{patch_fcn_comp}. On the other hand, FCNs are built from locally connected layers and do not contain fully connected layers. Hence, FCNs can perform training and inference on a per-pixel level efficiently~\cite{long2015fully}. A particular FCN known as U-Net shows competitive performance in image segmentation, especially in medical imaging but also in other domains such as robotics through the use of video data~\cite{shvets2018automatic}. U-Net consists of a series of convolutional layers that downsamples the image (encoder phase), 2 convolutional layers (bottleneck), and a series of convolutional layers that upsamples the image (decoder phase)~\cite{DBLP:journals/corr/RonnebergerFB15}. The downsampling path captures context while the symmetric upsampling path enables precise localization ~\cite{DBLP:journals/corr/RonnebergerFB15}.

There are generally two approaches in training a CNN, from scratch and transfer learning~\cite{garcia2018behavior}. Using the transfer learning approach, pretrained FCNs on ImageNet~\cite{deng2009imagenet} generally provide improved performance over training models from scratch. This approach is especially useful when limited data or computational resources is present~\cite{imagenet_cvpr09}. In particular, TernausNet, a U-Net with a VG11 encoder pretrained on ImageNet speeds up convergence and improves segmentation accuracy over the trained from scratch counterpart on boundary segmentation of cars~\cite{DBLP:journals/corr/abs-1801-05746}. 

Apart from the traditional method of using a single image as input and the ground truth segmented mask, approaches to incorporate additional relevant data sources show improved performance on image segmentation tasks. For instance, a recent approach using U-Nets to incorporate input images from multiple time frames as a means to incorporate temporal information provide more promising results over simply using a single image in the task of urban land use classification~\cite{mendili2020towards}. 

For our particular task, supervised machine learning and deep learning approaches using FCNs to delineate farm parcels perform better than classical computer vision methods and have produced state-of-the-art results~\cite{ doi:10.1080/01431161.2016.1278312, 8886377, MUSYOKA2018AUTOMATICDO}. The traditional methods such as edge-based detectors perform reasonably well on delineating boundaries of regularly-shaped farmland areas but fail in cases with more complex farmland shapes. Furthermore, many of these traditional methods do not delineate boundaries contextually, a relationship that can be utilized by deep learning methods~\cite{MUSYOKA2018AUTOMATICDO}.

% \burak{Not clear}

%-------------------------------------------------------------------------
\section{Problem Statement}
Given an input of satellite images represented with $x_{i}^{t} \in \mathcal{X}^{t}$ from an area with geo-coordinates $c_{i}=(lat_{i}, lon_{i})$, we output a binary mask represented with $\hat{y}_{i} \in \mathcal{\hat{Y}}$. Here, the index $t$ represents the images over time from the area centered at $c_i$. $t=0,t=1,$ and $t=2$ represent the time ranges Jan-March, April-June, and July-Sept respectively. The image captured in April-June is represented with $x_{i}^{t=1}$ and it aligns with ground truth mask $y_{i}$ temporally. In our study, we model the output mask to contain pixel wise binary labels. Specifically, our models output two different masks: (1) delineated boundaries between farm parcels, (2) segmentation masks of farm parcels. In the first case, the goal is to assign the pixel with image coordinates, $m$ and $n$, the label $1$ if it lies on the boundary of a farm parcel as $f_b:x_{i}^{t}(m,n) \mapsto \hat{y}_{i}(m, n)=1$ where $f_b$ represents the boundary segmentation network. In the second task, our goal is to map a pixel in the image space with coordinates $m$ and $n$ to label 1 if it lies inside the farm parcel as $f_a:x_{i}^{t}(m,n) \mapsto \hat{y}(m, n)=1$ where $f_a$ represents the area segmentation network.
% \burak{explain what t=2,3,... is}

% \begin{figure}[]
% \centering
% \begin{subfig}{.5\linewidth}
%   \centering
%   \includegraphics[width=.4\linewidth]{latex/spatial_unet.png}
%   \caption{Spatial U-Net}
%   \label{fig:spatial_unet}
% \end{subfig}%
% \begin{subfig}{.5\linewidth}
%   \centering
%   \includegraphics[width=.4\linewidth]{latex/spatio_temporal_unet.png}
%   \caption{Spatio-temporal U-Net}
%   \label{fig:spatio_temproal_unet}
% \end{subfig}
% \caption{Experimented models}
% \label{fig:unets}
% \end{figure}

% \begin{figure}[h!]
% \centering
% \includegraphics[width=1.0\linewidth]{latex/spatial_unet.png}
% \caption{Spatial U-Net}
% \label{fig:spatial_unet}
% \end{figure}

% \begin{figure}[h!]
% \centering
% \includegraphics[width=1.0\linewidth]{latex/spatio_temporal_unet.png}
% \caption{Spatio-temporal U-Net}
% \label{fig:spatial_unet}
% \end{figure}

\subsection{Dataset}

\begin{figure}[h]
  \centering
  \includegraphics[width=1.0\linewidth]{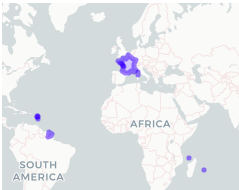}
  \caption{Spatial coverage of the shapefile used for the dataset to sample polygons.}
%   \burak{Remove the Spatial Coverage Part at the top of the image.}
\label{fig:shapefile}
\end{figure}

\subsubsection {Data Description}
The dataset represented with $(\mathcal{X}^{t}, \mathcal{Y})$ consists of Sentinel-2 satellite imagery ($224px \times 224px $ RGB image corresponding to $2.24km \times 2.24km$ of land area) along with corresponding binary masks of boundaries and areas of farm parcels~\cite{sentinel2}. Sample instances from the dataset are shown in Fig.~\ref{fig:example_data_instance}. Sentinel-2 is used over other satellite imaging datasets such as DigitalGlobe as Sentinel-2 covers a much larger area of coverage per image, allowing more farm parcels to be delineated in a single image. Furthermore, Sentinel-2 is freely available and has a relatively short revisit time globally, making the dataset a suitable choice in collecting the most recent cadastral information through satellite imaging.

\begin{figure}[h]
    \centering
    \subfloat[Example polygon from the shapefile ]{{\includegraphics[width=0.43\linewidth]{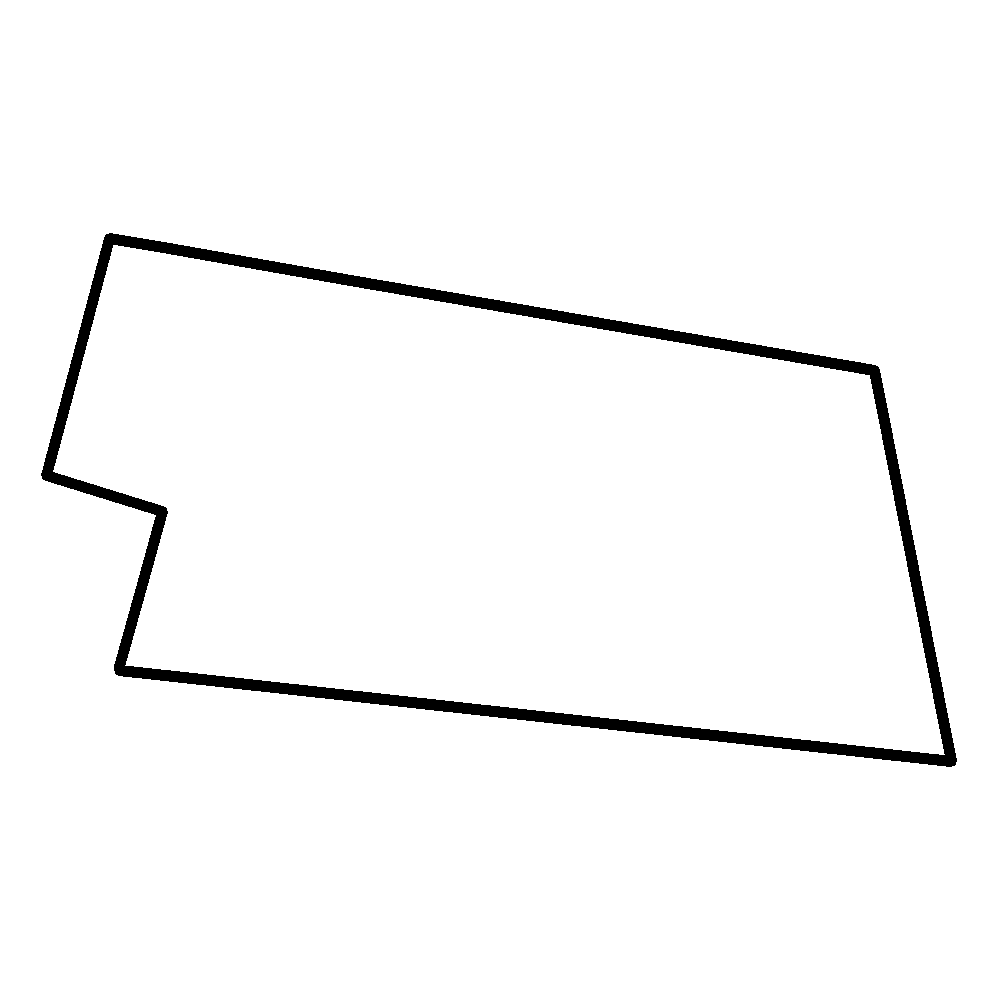} }}%
    \qquad
    \subfloat[Region where the polygon is overlaid. ]{{\includegraphics[width=0.43\linewidth]{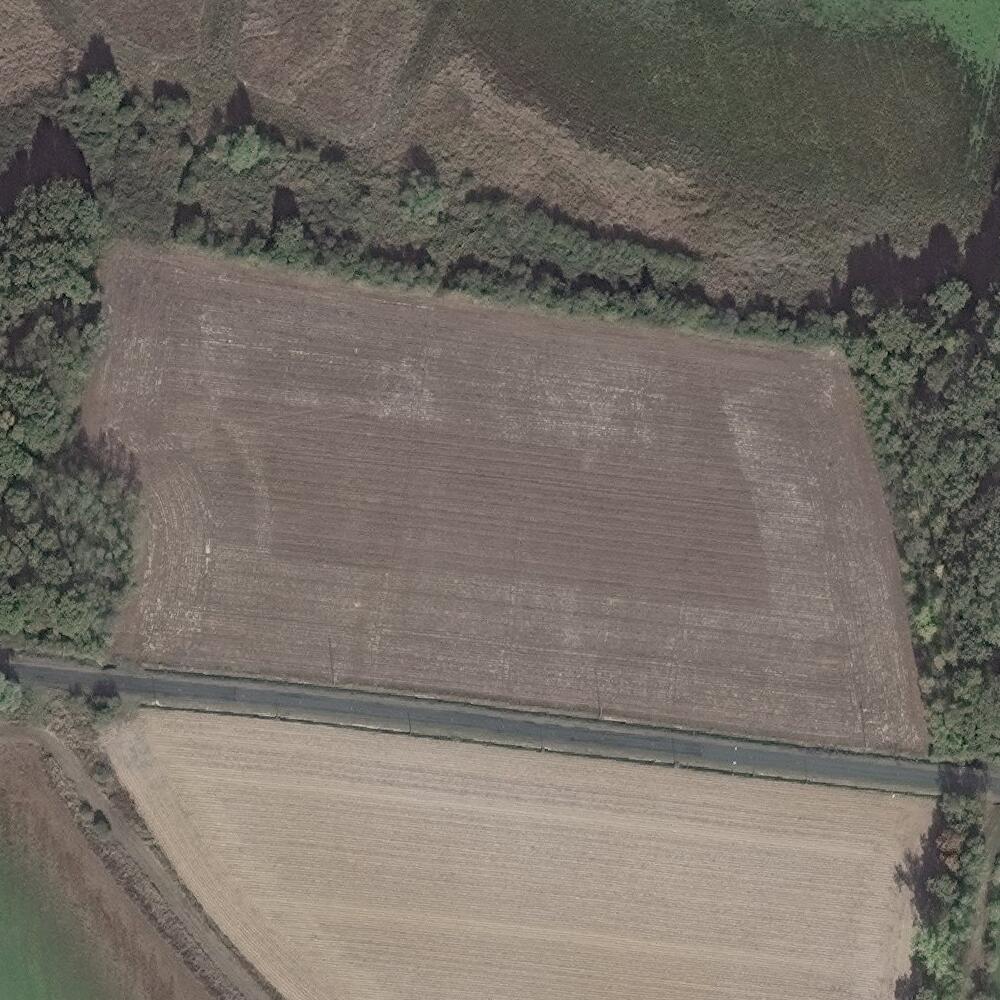} }}%
    \caption{Sample polygon from the shapefile and the corresponding region to be overlaid. The satellite image is zoomed-in to better visualize where the polygon would be overlaid.}%
    \label{fig:sample_polygon}%
\end{figure}

\begin{figure*}[h!]
    \centering
    \subfloat[Spatial U-Net]{{\includegraphics[width=0.47\textwidth]{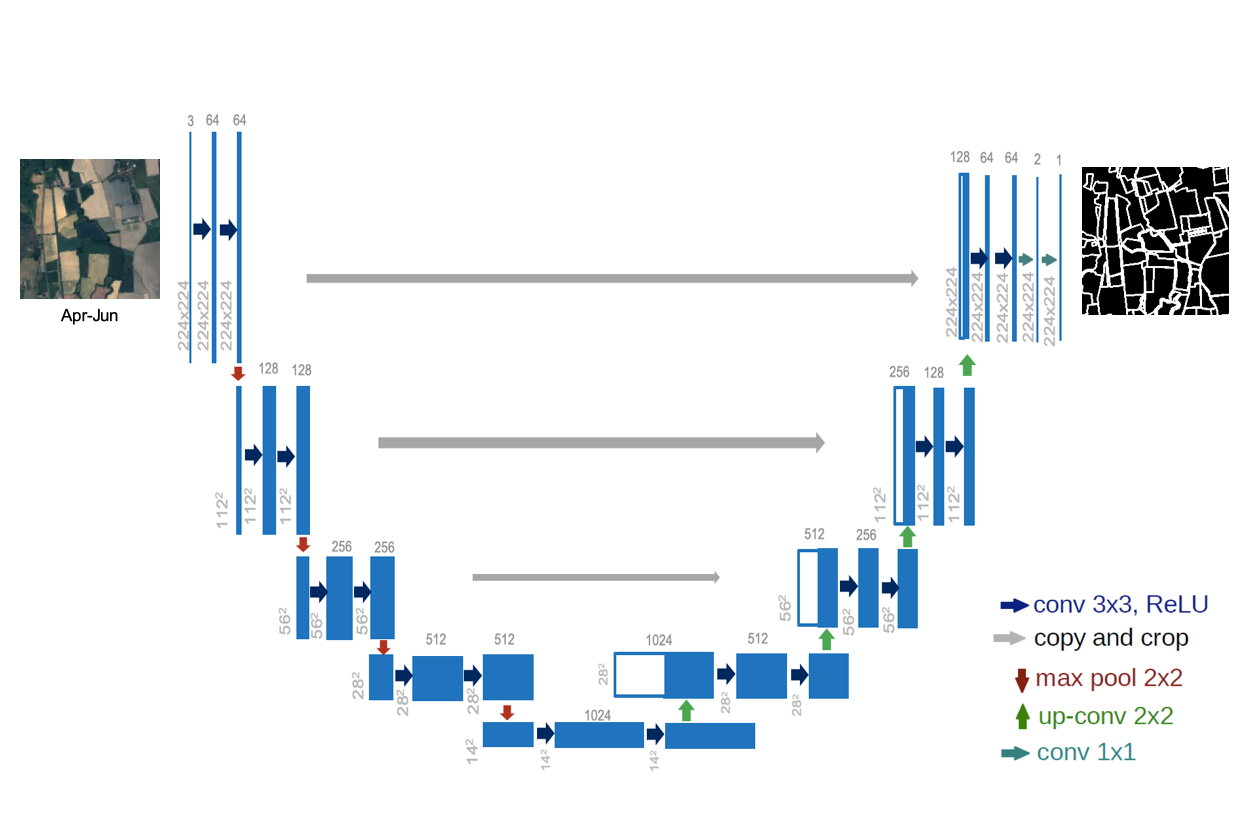} }}%
    \qquad
    \subfloat[Spatio-temporal U-Net]{{\includegraphics[width=0.47\textwidth]{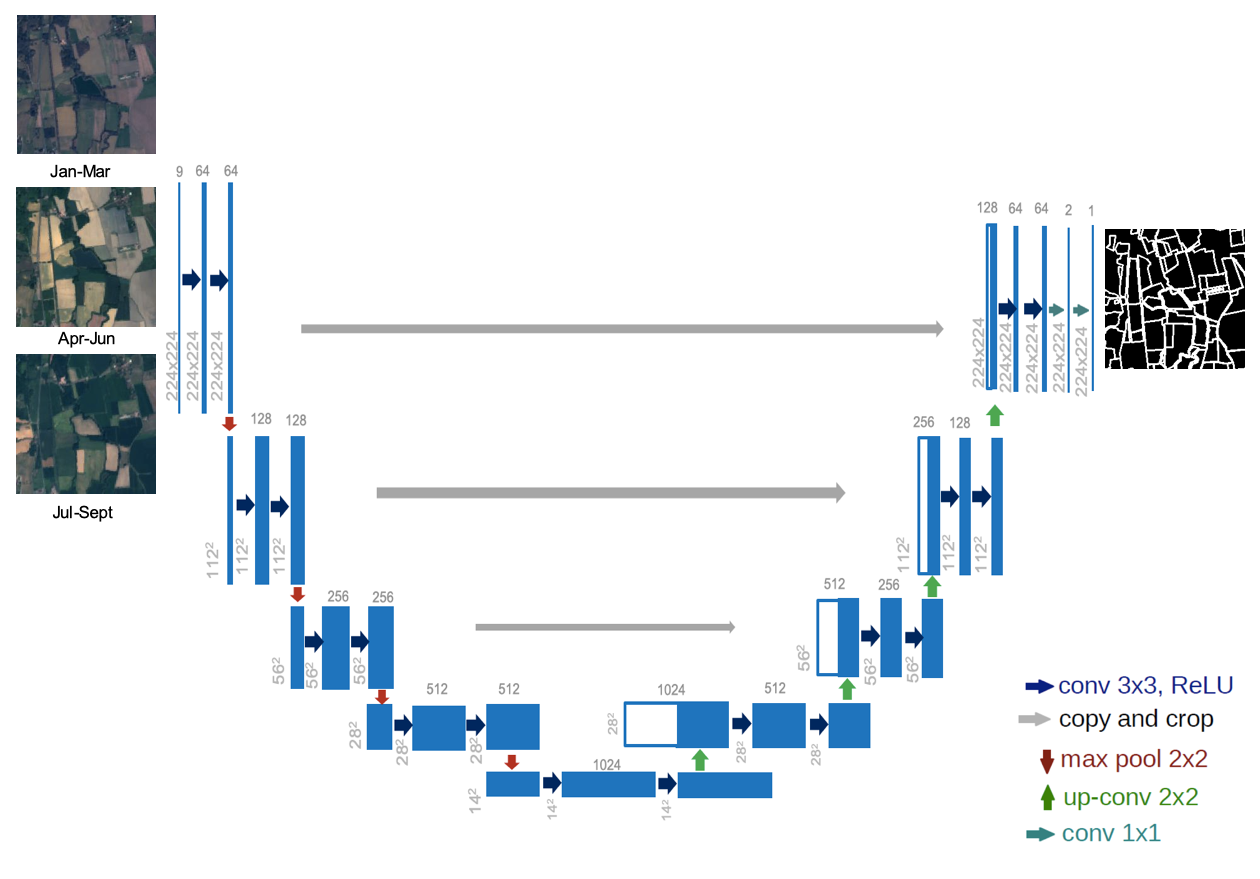} }}%
    \caption{U-Net models (a type of Fully Convolutional Network) used for our boundary and area segmentation task. The Spatial U-Net takes in a single satellite image as input and outputs the binary mask whereas the Spatio-temporal U-Net takes into account 3 images from different timestamps to output the binary mask.}%
    %\burak{Explain what these models are briefly in here.}
    \label{fig:models}%
\end{figure*}

\subsubsection {Data Generation}
First, we started with downloading the shapefile consisting of the polygons (in Lambert-93 coordinate system) of farm parcel boundaries of France (2017) with the spatial coverage shown in Fig.~\ref{fig:shapefile} from a publicly available resource~\footnote[1]{https://www.data.gouv.fr/en/datasets/registre-parcellaire-graphique-rpg-contours-des-parcelles-et-ilots-culturaux-et-leur-groupe-de-cultures-majoritaire/}. Fig.~\ref{fig:sample_polygon} presents a sample polygon from the shapefile. Afterwards, we projected the existing coordinate system of all the polygons to the geographic coordinate system of latitude/longitude. We sampled 2000 regions by randomly sampling 2000 $c_i (lat_i, lon_i)$ that have at least one farm parcel in its vicinity (within 2.24km radius) in France. Our image geolocation parameter $c_i (lat_i, lon_i)$ represents the center of the square region $i$ that spans $2.24km \times 2.24km$. To ensure that there is at least one farm parcel in each region $i$, we sampled random polygons in the downloaded shapefile and calculated the centroid (in latitude and longitude) of each polygon until we collected 2000 centroids that do not overlap. We then selected polygons in the original shapefile that are within the bounds of these 2000 square regions to create a filtered shapefile. To generate the binary boundary mask for each square region $i$, we parsed the filtered shapefile to combine the set of polygons of farmland boundaries that are only relevant for that particular square region $i$. Afterwards, we projected the geographic coordinate system $(lat_i, lon_i)$ onto the $224px \times 224px$ image based on linear-scaling. To draw the boundaries, we used OpenCV's polylines functionality~\cite{opencv_library}. We set the thickness of the boundary to be 2 pixels. For the corresponding binary mask of farmland areas, we used OpenCV's fillPoly functionality~\cite{opencv_library}. 

Second, we collected the Sentinel-2 RGB satellite imagery $x_i^t$ ($224px \times 224px$) corresponding to each square region $i$ over 3 time ranges $t= 0$ (January - March), $t = 1$ (April - June), and $t = 2$ (July-September) in 2017. Each $x_i^t$ is a composite satellite image over the 3-month time range. 

% \begin{figure}[h!]
%   \centering
%   \includegraphics[width=1.0\linewidth]{./polygon_example.png}
%   \caption{Example polygon from the shapefile}
%   \burak{Would be nice to show the image where it represents.}
% \label{fig:shapefile}
% \end{figure}

\section{Methods}
Given our dataset represented with $(\mathcal{X}^{t}, \mathcal{Y})$ we propose two methods to segment farmland boundaries and areas. Our first method uses only one image that is representative of time range (April-June), $x_{i}^{t=1}$. The second method, on the other hand, utilizes 3 images, $x_{i}^{t=0,1,2}$, that is representative of 3 time ranges mentioned previously. 

\begin{figure*}[!h]
    \centering
    {\includegraphics[width=0.8\textwidth]{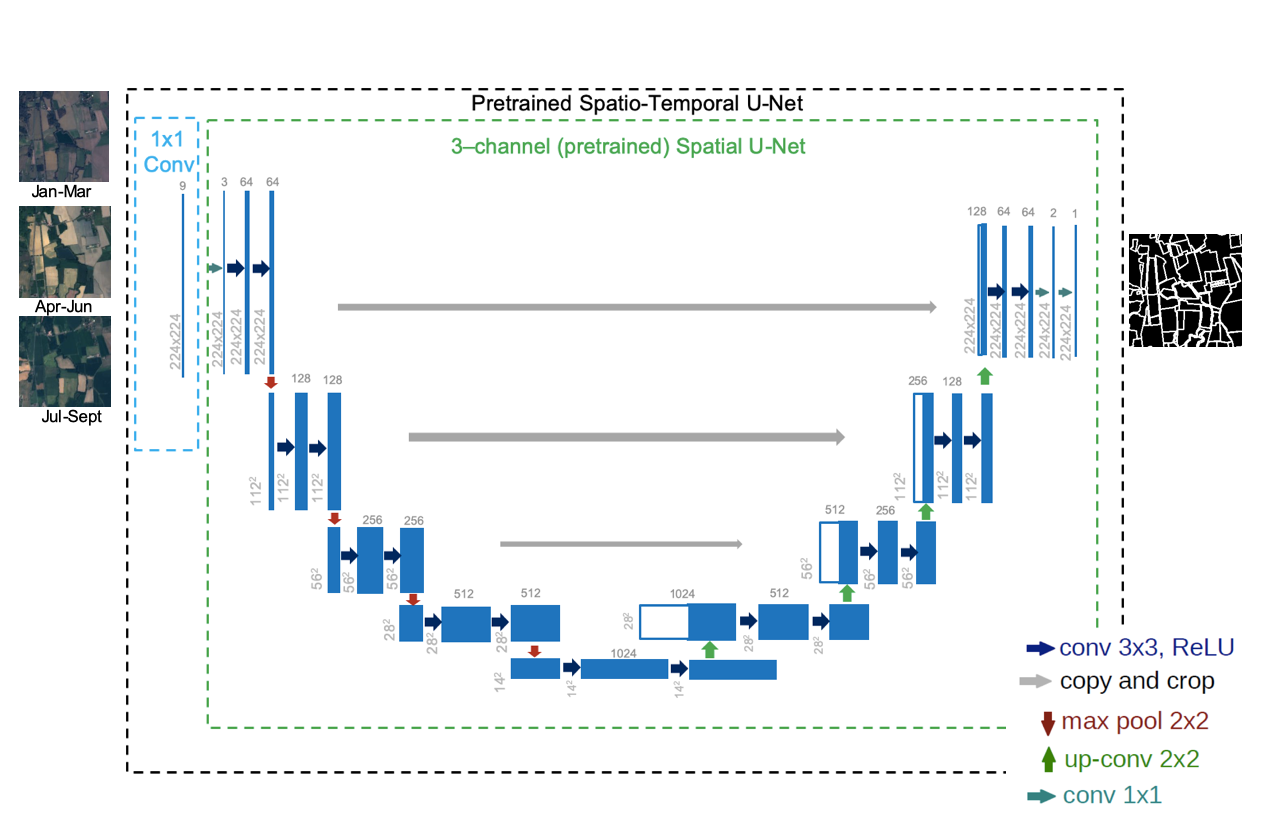} }%
    \caption{Pretrained Spatio-temporal U-Net. The pretrained Spatio-temporal U-Net differs from the (pretrained) Spatial  U-Net (and non-pretrained Spatio-temporal U-Net) in that this model adds an additional convolution layer to map the 9 channels from 3 RGB satellite images into the 3-channel pretrained Spatial U-Net.}%
    %\burak{Why do we need this figure? - this pretrained model is different from the model trained from scratch}
    \label{fig:pretrained_model}%
\end{figure*}

\subsection{Spatial U-Net}
We decided to use U-Net as it has shown competitive performance over multiple domains in image segmentation~\cite{DBLP:journals/corr/RonnebergerFB15}. We coin our U-Net model~\cite{DBLP:journals/corr/RonnebergerFB15} for this task as the \emph{Spatial U-Net} which includes 3 channels for the input. The model takes in a single RGB satellite image $x_i^1$ (April-June) as input and outputs a binary mask of boundaries/areas. We experiment with both a Spatial U-Net trained from scratch and also a pretrained Spatial U-Net whose encoder is initialized with weights learned on the ImageNet classification task~\cite{imagenet_cvpr09}. 

\subsection{Spatio-temporal U-Net}
In addition to the Spatial U-Net, we propose a \emph{Spatio-temporal U-Net} model that utilizes 9 channels representing 3 RGB images over time as $(t=0, t=1, t=2)$ from the area $i$ centered at $c_i$. In some cases we found that an image out of 3 images is not available. In such cases, we replace the missing image with $x_i^1$, which always exists.

We also experiment with a pretrained Spatio-temporal U-Net. To use the pretrained weights, the architecture of the pretrained Spatio-temporal U-Net instead includes a convolution layer that simply maps the 9 channels into the 3-channel pretrained Spatial U-Net as shown in Fig.~\ref{fig:pretrained_model}. 

% \burak{Can you find a citation for this? - Hmm, I can find a citation for 1x1 convs but not sure exactly for this}
\section{Experiments}

% \begin{figure*}[h!]
%     \centering
%     \subfloat[Pretrained Spatio-temporal U-Net]{{\includegraphics[width=0.47\textwidth]{./pretrained_saptio_temporal.png} }}%
%     \qquad
%     \subfloat[Pretrained Spatial U-Net ]{{\includegraphics[width=0.47\textwidth]{./spatial_unet.png} }}%
%     \caption{The pretrained Spatio-temporal U-Net differs from the pretrained Spatial U-Net (and the vanilla Spatio-temporal U-Net) in that this model adds an additional convolution layer to map 9 channels from three RGB satellite images into the 3-channel pretrained Spatial U-Net.}%
%     %\burak{Explain what these models are briefly in here.}
%     \label{fig:models}%
% \end{figure*}

\begin{table*}[h!]
\resizebox{0.98\linewidth}{!}{
\begin{tabular}{@{}ccccc@{}}
\toprule
 & \textbf{\begin{tabular}[c]{@{}c@{}}Spatial\\ U-Net \end{tabular}} & \textbf{\begin{tabular}[c]{@{}c@{}}U-Net\\ (\emph{Pretrained on ImageNet }) \end{tabular}} & \textbf{\begin{tabular}[c]{@{}c@{}}Spatio-temporal\\ U-Net \end{tabular}} & \textbf{\begin{tabular}[c]{@{}c@{}}Spatio-temporal U-Net\\ (\emph{Pretrained on ImageNet})\end{tabular}} \\ \midrule
\textbf{Dice Score - \emph{Boundary}} & 0.56 & 0.54 & 0.60 & \textbf{0.61} \\
\textbf{Dice Score - \emph{Area}} & 0.72 & 0.75 & 0.80 & \textbf{0.81} \\
\textbf{Accuracy - \emph{Boundary}} & 0.76 & 0.78 & 0.81 & \textbf{0.82} \\
\textbf{Accuracy - \emph{Area}} & 0.71 & 0.77 & 0.82 & \textbf{0.83} \\ \bottomrule
\end{tabular}}
\caption{Boundary and area segmentation results in terms of the Dice score and accuracy.}
\label{tbl:result}%
\end{table*}

\subsection{Metrics}
We evaluated the models on the Dice score and per-pixel accuracy. The Dice score is also known as the F1 score and formulated as
\begin{equation}
DICE = \frac{2TP}{2TP + FP + FN}
\end{equation}
where TP = True Positive, FP = False Positive and FN = False Negative.

While accuracy is an intuitive metric, the Dice score is considered as a better metric in the cases with class imbalance between boundary/non-boundary pixels. It has been widely used in image segmentation tasks across different domains ~\cite{zou2004statistical,  DBLP:journals/corr/KamnitsasLNSKMR16, sheehan2018learning}.

\subsection{Implementation Details}
We implemented all the U-Net models in Keras~\cite{chollet2015keras}. The U-Net models that are not pretrained incorporates dilated convolutions for the encoder instead of regular convolutions~\cite{yu2015multi} as dilated convolutions can better aggregate the context of the whole image. We initialized the weights of the encoder of pretrained models from the ImageNet classification task with the ResNet-34 backbone by extracting the same number of layers from the ResNet-34 as the number of encoder layers in our models~\cite{Yakubovskiy:2019, imagenet_cvpr09}. We trained all the models using the Adam optimizer~\cite{adam}, a learning rate of 1e-4, batch size of 6 over 200 epochs with NVIDIA GeForce GTX 1080 Ti. Finally, we experimented with both the binary cross entropy and the Dice loss functions to train the models. 

For experiments with the Spatial U-Net, each data instance is represented by $(x_i^1, y_i)$, which corresponds to the satellite image and ground truth mask collected in April-June. As for experiments with the Spatio-temporal U-Net, each data instance is represented by $(\{x_i^0, x_i^1, x_i^2\}, y_i)$. The dataset is split using a random distribution and the split percentages for train/validation/test are 80/10/10, respectively. The implementation of the whole pipeline can be found in our github repository~\footnote[2]{Code repo: https://github.com/sustainlab-group/ParcelDelineation}.

\subsection{Results}
We provide the Dice score and per-pixel accuracy for both boundary and area segmentation tasks for all the models trained using the binary cross entropy loss in Table~\ref{tbl:result}. We also provide qualitative results (i.e. samples of predictions of output binary masks) for the pretrained Spatio-temporal U-Net in Fig.~\ref{fig:example_prediction} and a qualitative comparison between the pretrained Spatio-temporal U-Net and pretrained Spatial U-Net in Fig.~\ref{fig:spatio_temporal_spatial_qual_comp}.

As shown in Table~\ref{tbl:result}, the Spatial U-Net performs the worst on the area segmentation task with a Dice score of \textbf{0.72} and accuracy of \textbf{0.71}. The pretrained Spatial U-Net performs the worst on the boundary segmentation task with a Dice score of \textbf{0.54} though the pretrained Spatial U-Net does have a higher accuracy of \textbf{0.75} compared to the Spatial U-Net's accuracy of \textbf{0.72}.

The pretrained Spatio-temporal U-Net performs best for both boundary and area segmentation tasks on all metrics, achieving \textbf{0.61} Dice score and \textbf{0.82} accuracy on the boundary segmentation task and \textbf{0.81} Dice score and \textbf{0.83} accuracy on the area segmentation task.

\section{Analysis}
\subsection{Quantitative Analysis}
% \burak{This is an important part and I did not understand much. Can you use shorter sentences.}
\paragraph{Area vs. Boundary Segmentation} The area segmentation results seem to indicate a very powerful performance from the models. However, a large region of each satellite image is farmland area. Therefore, the quantitative border segmentation results provide a more indicative measure of the performance of each model. Hence, we put most of the emphasis on quantitative analysis on the boundary segmentation task. However, the area segmentation results are still useful for qualitative comparisons. 

\paragraph{Pretrained vs. Trained From Scratch Spatial U-Net} Despite limited amount of training data, we observe from Table~\ref{tbl:result} that even the baseline Spatial U-Net performs reasonably well with a Dice score of \textbf{0.56} and \textbf{0.72} on the boundary and area segmentation tasks respectively. However, we are surprised that the pretrained Spatial U-Net has a lower Dice score for the boundary segmentation task (though the accuracy is higher for the pretrained Spatial U-Net). One main difference between the pretrained and trained from scratch Spatial U-Net is the use of dilated convolutions in the latter model. The dilated convolutions are found to be effective in improving the learned contextual information in satellite images~\cite{hamaguchi2018effective,zhou2018d}. 

\paragraph{Spatio-temporal vs. Spatial U-Net} When we compare the Spatio-temporal U-Net with the Spatial U-Net, we observe an increase in performance on the boundary segmentation task with the Spatio-temporal U-Net. This improvement indicates the utility of temporal information. The Spatio-temporal U-Net has an increase of 7\% in Dice score and an increase of 11\% in accuracy over the Spatial U-Net. Across the pretrained counterparts, the pretrained Spatio-temporal U-Net also has an increase of 13\% in Dice score and an increase of 5\% in accuracy over the pretrained Spatial U-Net. 

On the other hand, pretraining seems to only slightly boost the performance of the model though it does greatly increase the convergence rate during training. The effects of pretraining can not be completely isolated, however, as the architecture of the pretrained and the non-pretrained counterparts slightly differ for the Spatio-temporal U-Net models.

\begin{figure*}[t!]%
\centering
\begin{tabular}{c}
\subfloat[Input (Apr-Jun, Jul-Sept)]{
{\includegraphics[width=0.152\linewidth]{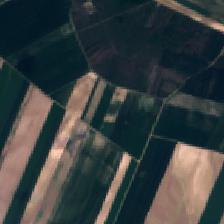} },
{\includegraphics[width=0.152\linewidth]{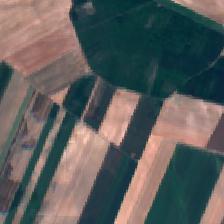}}}
\subfloat[Border prediction] {
{\includegraphics[width=0.152\linewidth]{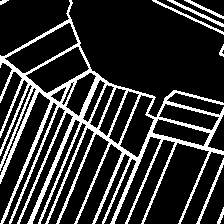} },
{\includegraphics[width=0.152\linewidth]{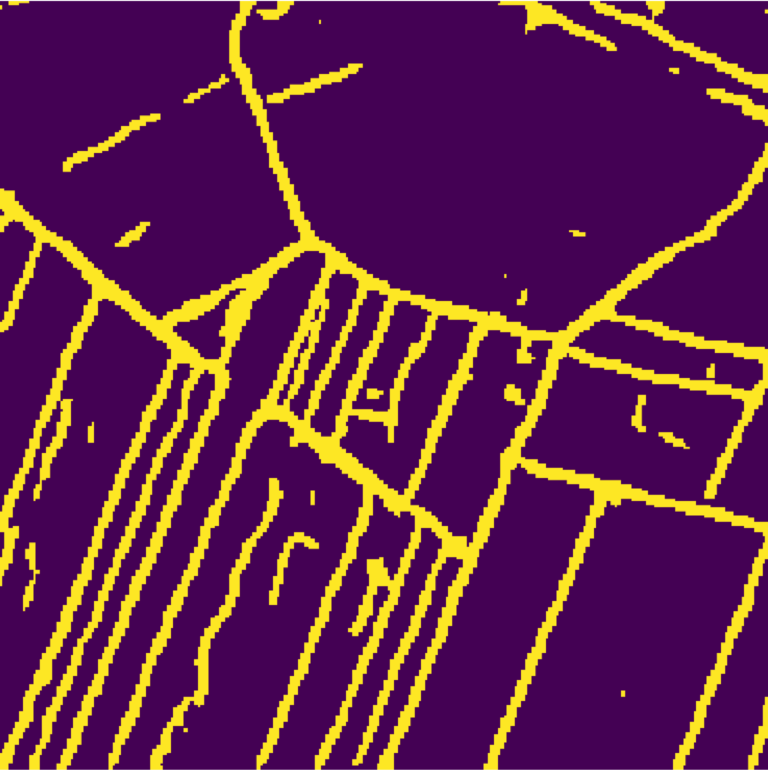} }}
\subfloat[Area prediction]{
{\includegraphics[width=0.152\linewidth]{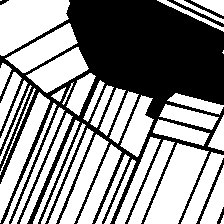} },
{\includegraphics[width=0.152\linewidth]{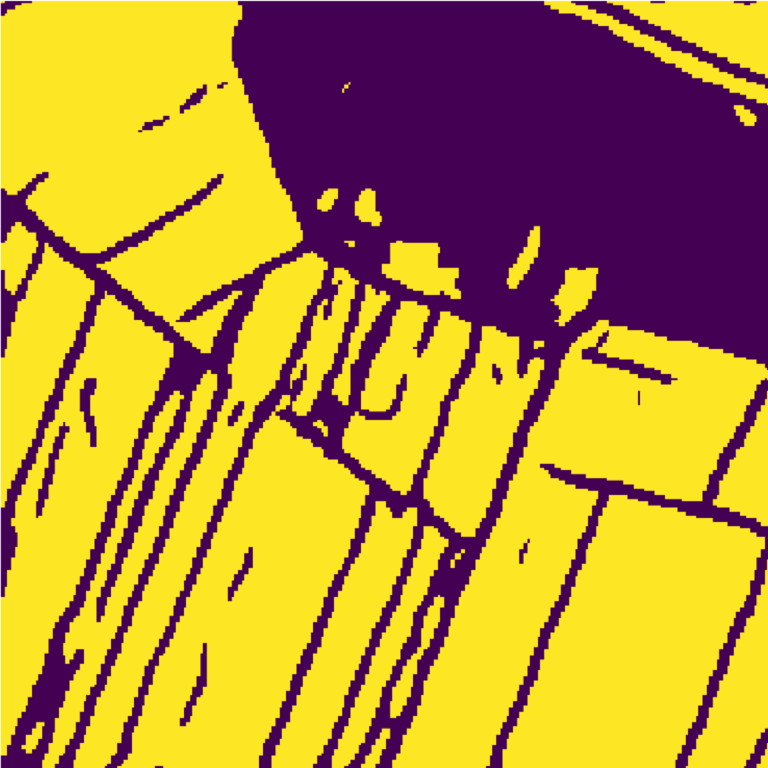} }}  \\
\subfloat[Input (Jan-Mar, Apr-Jun, Jul-Sept)]{
{\includegraphics[width=0.128\linewidth]{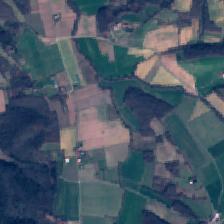} },
{\includegraphics[width=0.128\linewidth]{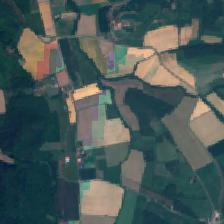} },
{\includegraphics[width=0.128\linewidth]{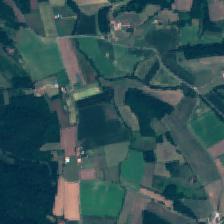} }}
\subfloat[Border prediction] {
{\includegraphics[width=0.128\linewidth]{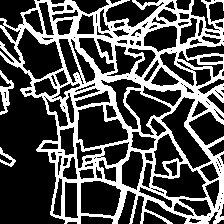} },
{\includegraphics[width=0.128\linewidth]{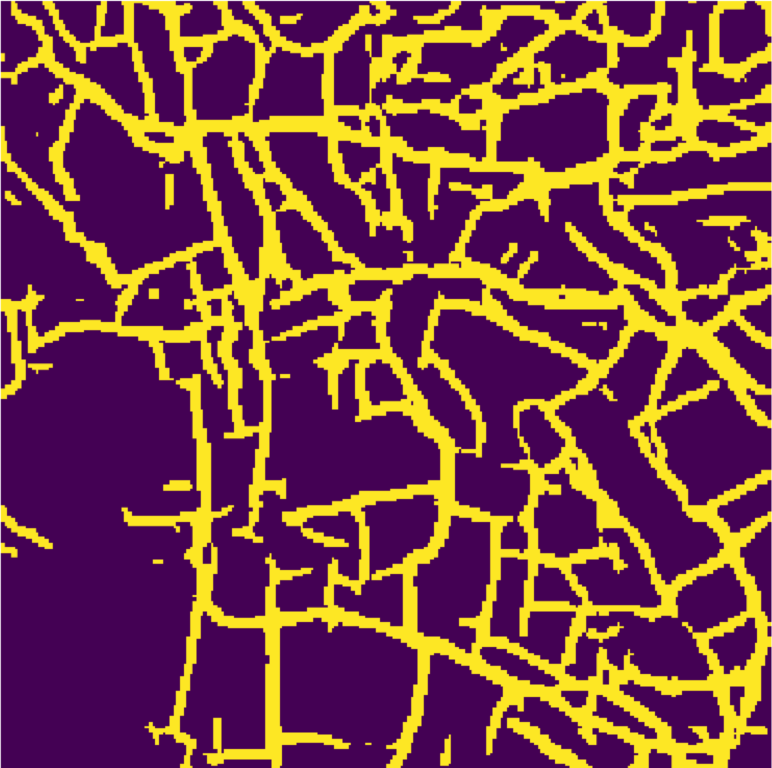} }}
\subfloat[Area prediction]{
{\includegraphics[width=0.128\linewidth]{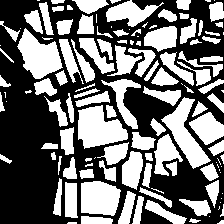} },
{\includegraphics[width=0.128\linewidth]{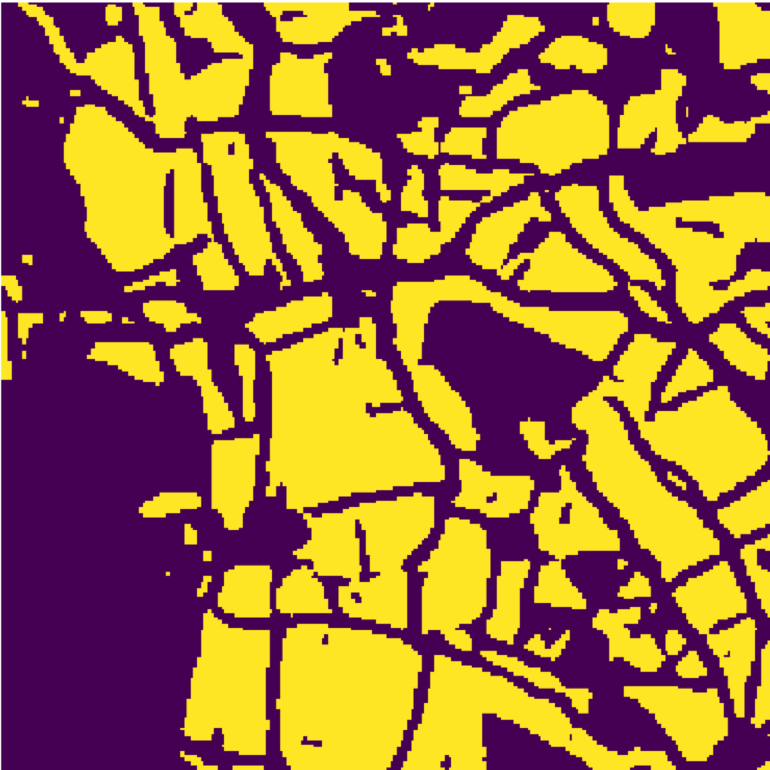} }}  \\
\subfloat[Input (Apr-Jun, Jul-Sept)]{
{\includegraphics[width=0.152\linewidth]{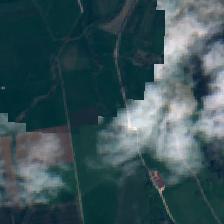} },
{\includegraphics[width=0.152\linewidth]{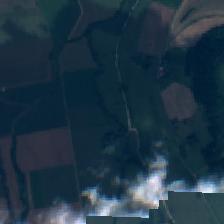} }}
\subfloat[Border prediction] {
{\includegraphics[width=0.152\linewidth]{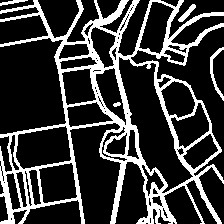} },
{\includegraphics[width=0.152\linewidth]{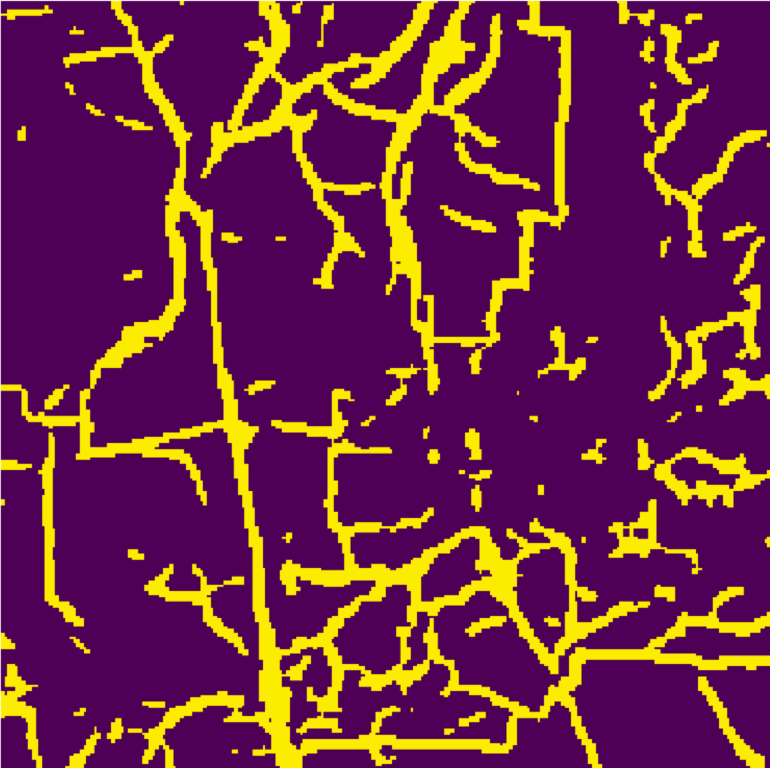} }}
\subfloat[Area prediction]{
{\includegraphics[width=0.152\linewidth]{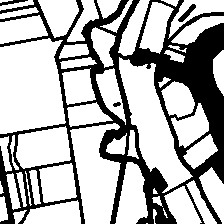} },
{\includegraphics[width=0.152\linewidth]{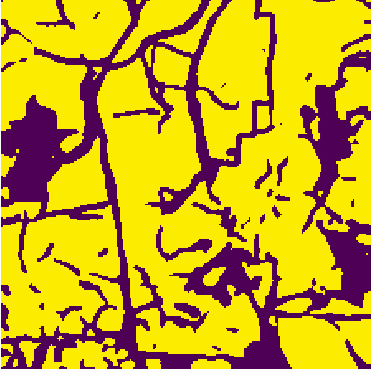} }}  
\\
\subfloat[Input (Apr-Jun, Jul-Sept)]{
{\includegraphics[width=0.152\linewidth]{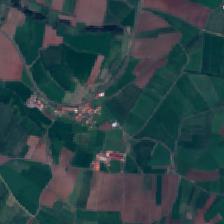} },
{\includegraphics[width=0.152\linewidth]{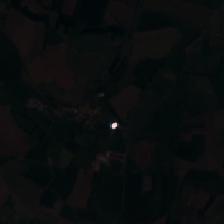} }}
\subfloat[Border prediction] {
{\includegraphics[width=0.152\linewidth]{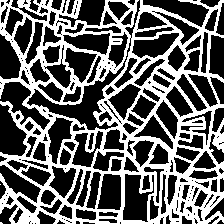} },
{\includegraphics[width=0.152\linewidth]{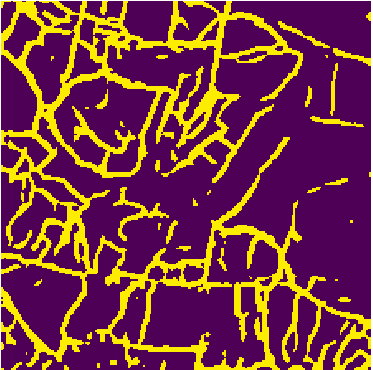} }}
\subfloat[Area prediction]{
{\includegraphics[width=0.152\linewidth]{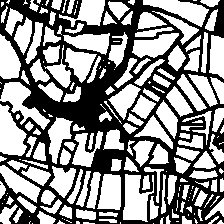} },
{\includegraphics[width=0.152\linewidth]{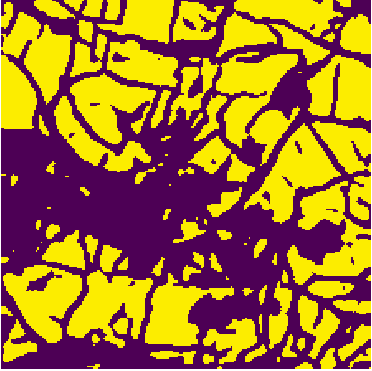} }} 
\end{tabular}
    \caption{Example predictions by the Spatio-temporal U-Net. From left to right, the first 2 or 3 images (depending on availability of Jan-Mar image) are the input satellite images. The next set of two images are the ground truth binary mask and the model's prediction respectively for the border segmentation task. The final set of two images are the ground truth binary mask and the model's prediction respectively for the area segmentation task.}%
    % \burak{Can you put the original image and mask to the left before the mask overlaied image?}
    \label{fig:example_prediction}%
\end{figure*}

\subsection{Qualitative Analysis}

\begin{figure}[h!]%
\centering
\begin{tabular}{c}
\subfloat[Input satellite images (Apr-Jun on the left and Jul-Sept on the right)] {
{\includegraphics[width=0.476\linewidth]{orig_apr_jun_3.jpeg} },
{\includegraphics[width=0.476\linewidth]{orig_jul_sept_3.jpeg} }
}\\
\subfloat[Border predictions by the pretrained Spatio-temporal U-Net, pretrained Spatial U-Net and ground truth]{
{{\includegraphics[width=0.312\linewidth]{./predict_3.png} } {\includegraphics[width=0.312\linewidth]{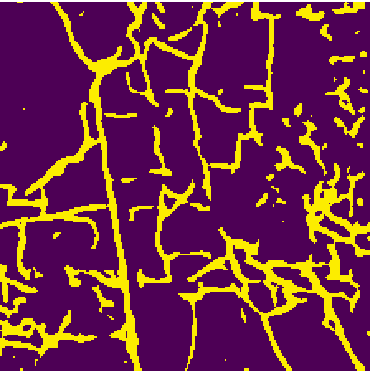} }
{\includegraphics[width=0.312\linewidth]{./orig_mask_3.png}}}}\\
\subfloat[Area predictions by the pretrained Spatio-temporal U-Net, pretrained Spatial U-Net and ground truth]{
{{\includegraphics[width=0.312\linewidth]{./predict_area_3.png} }
{\includegraphics[width=0.312\linewidth]{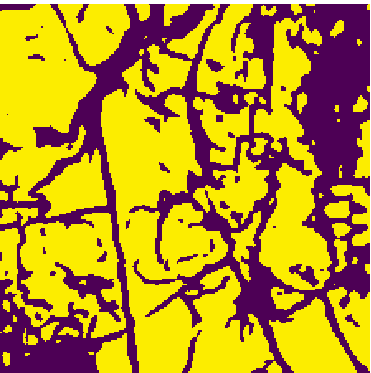} }
{\includegraphics[width=0.312\linewidth]{./orig_mask_filled_3.png} }}} % \\ 
% \subfloat[]{{\includegraphics[width=1cm]{jan_march_4.jpeg} }, {\includegraphics[width=1cm]{apr_jun_4.jpeg} }, {\includegraphics[width=1cm]{jul_sept_4.jpeg} },
% {\includegraphics[width=1cm]{out_4.png} },
% {\includegraphics[width=1cm]{out_4_filled.png} }
\end{tabular}
    \caption{Border and area segmentation results comparison between the pretrained Spatial U-Net and pretrained Spatio-temporal U-Net. Row (a) presents satellite images with cloud coverage. Row (b) and (c) are border and area predictions from the two compared models respectively. For row (b) and row (c), from left to right, the first image is the prediction from the pretrained Spatial U-Net; the second image is the prediction from the pretrained Spatio-temporal U-Net; the third image is the ground truth.}%
    \label{fig:spatio_temporal_spatial_qual_comp}%
\end{figure}

\subsubsection{Qualitative performance of the Spatio-temporal U-Net}

Fig.~\ref{fig:example_prediction} presents samples of predictions using our pretrained Spatio-temporal model. We observe that the pretrained Spatio-temporal model provides predictions reasonably well on the regularly shaped farmland areas as shown in Fig.~\ref{fig:example_prediction} (a)-(c) for both the border and area segmentation tasks. However, the model misses more positive predictions of farmland pixels for irregularly shaped and densely packed farmland areas as shown in Fig.~\ref{fig:example_prediction} (d)- (f) and (j)-(l), it still provides reasonable predictions. In particular, for the area segmentation task, we observe that the model is often able to predict areas that may seem like farmland areas at a cursory glance to the human eye but are instead buildings/barren areas/bushes correctly. 

\subsubsection{Cloud-covered areas}

In cases where the input image in a particular timestamp $t$ is obscured by clouds (as in Fig.~\ref{fig:example_prediction} input image (g)), the pretrained Spatio-temporal model performs worse compared to cases of non-obscured input images. However, the model still provides correct predictions in certain cloud-covered regions. This may be because even though the rest of the image(s) may have slightly different farmland shapes, they still help `fill in the gap' for the cloud-obscured image. We notice that the model still segments some farmland areas and boundaries correctly despite image(s) being obscured by the clouds.

Fig.~\ref{fig:spatio_temporal_spatial_qual_comp} provides a more concrete comparison between the pretrained Spatio-temporal U-Net and the pretrained Spatial U-Net for the same cloud-covered input satellite images presented in Fig.~\ref{fig:example_prediction} (g). As observed for both border and area segmentation predictions in Fig.~\ref{fig:spatio_temporal_spatial_qual_comp}, the pretrained Spatial U-Net misses positive predictions of boundaries and areas in regions where clouds are present (i.e. on the upper right corner of the satellite image in April-June in Fig.~\ref{fig:spatio_temporal_spatial_qual_comp} (a)). However, the pretrained Spatio-temporal model is able to `fill in the gap' using images from two different time ranges (April-June and July-Sept). 

For boundary segmentation, the pretrained Spatio-temporal model outputs smoother and more connected boundaries in these cloud-covered areas compared to the pretrained Spatial U-Net. For area segmentation, the pretrained Spatio-temporal U-Net correctly predicts a large part of the cloud-covered area as farmland pixels whereas the pretrained Spatial U-Net mostly predicts cloud-covered areas to be non-farmland area pixels. This further validates the hypothesis of Spatio-temporal models being able to `fill in the gap'. To better handle clouds, we can use a cloud removal model as proposed in~\cite{sarukkai2020cloud} that utilizes 3 Sentinel-2 images over time in an area to generate cloud-free image of the same area. Similarly to~\cite{sarukkai2020cloud}, we use 3 Sentinel-2 images over time and we can design a multi-task model that segments parcel boundary/areas and generates a cloud-free image jointly.

\subsubsection{Error Analysis}
While the pretrained Spatio-temporal performs reasonably well across different shapes of farmland areas, the model mislabels pixels for farm boundaries/areas as non-farm boundaries/areas in certain cases. Specifically, for the boundary segmentation task, the model often predicts `broken' boundaries that should instead be connected. We notice a similar phenomenon for the area segmentation task in that the model incorrectly predicts smaller regions inside the farmland area as non-farmland pixels. We observe many of these errors in inputs where densely packed farmland regions and cloud-covered areas are present as shown in Fig. \ref{fig:example_prediction} (g)-(i) and Fig. \ref{fig:example_prediction} (j)-(l). One reason for these errors may have been due to the limitation of our dataset. An example of this could be occasional misalignment of polygons from the public dataset with the corresponding Sentinel-2 satellite image. This slight misalignment arises since we use 3-month composites for the Sentinel-2 images and a simple linear-based coordinate projection. 

Furthermore, in Fig.~\ref{fig:example_prediction} (j)-(k), where the different farmland regions have very similar colors, the model is not able to delineate boundaries and segment areas well, especially for smaller farmland areas. This error most likely arises from the limitation of using only RGB bands from the satellite images. 

\section{Conclusion}
In this study, we proposed the use of deep learning and open source datasets to segment farm parcel areas and boundaries in satellite images. In particular, we trained variants of the U-Net model on the Sentinel-2 images given the corresponding area/boundary masks.
We showed that the proposed Spatio-temporal U-Net achieves $83\%$ Dice score outperforming all Spatial-only models by around $3-5\%$. This shows that the additional temporal data can better highlight the faint boundaries of farmland parcels. As a future work, to improve predictions, we will experiment with other variations of U-Nets and models along with different transfer learning approaches. To analyze more extensively the impact of temporal information, we hope to experiment with different levels of granularity of temporal information. Finally, to further validate our results, we will apply our models on other regions that may have vastly different agricultural regions using a cross-spatial split.

The proposed use of deep learning and open source datasets provides a further step towards efficient and cheap automated cadastral data collection to provide faster and more accurate land policies decisions on agriculture, and climate change mitigation and adaptation.

{\small
\bibliographystyle{ieee_fullname}
\bibliography{egbib}
}

\end{document}